\documentclass[iop]{emulateapj}
\pdfoutput=1
\usepackage{graphicx, subfigure}
\usepackage{morefloats}
\usepackage{amsmath}
\usepackage{amssymb}
\usepackage{mathrsfs,amssymb}
\usepackage{url}
\newcommand{\degree}{\ensuremath{^\circ}}
\newcommand{\chandra}{\mbox{\it Chandra}}	 
\newcommand{\swift}{{\it Swift}}

\usepackage{mwe,tikz}
\usepackage[percent]{overpic}%

\begin{document}

\title{An Analysis of Chandra Deep Follow-up GRBs: Implications for Off-Axis Jets}

\author{Bin-Bin Zhang\altaffilmark{1,2,3,*}, Hendrik van Eerten\altaffilmark{4,5}, David N. Burrows\altaffilmark{1}, Geoffrey Scott Ryan\altaffilmark{5}, Philip A. Evans\altaffilmark{6}, Judith L. Racusin\altaffilmark{7}, Eleonora Troja\altaffilmark{7,8}, Andrew MacFadyen\altaffilmark{5}}

\altaffiltext{*}{Contact email: binbin.zhang@uah.edu}
\affil{$^{1}$Department of
Astronomy and Astrophysics, The Pennsylvania State University,
University Park, PA 16802, USA} 
\affil{$^{2}$Center for Space Plasma and Aeronomic Research (CSPAR), University
of Alabama in Huntsville, Huntsville, AL 35899, USA} 
\affil{$^{3}$Instituto de Astrof\'isica de Andaluc\'a (IAA-CSIC), P.O. Box 03004, E-18080 Granada, Spain}
\affil{$^{4}$Alexander von Humboldt Fellow, Max-Planck Institute for Extraterrestrial Physics (MPE), Postfach 1312, 85741 Garching, Germany}
\affil{$^{5}$Center for Cosmology and Particle Physics, Physics Department, New York University, New York, NY 10003}
\affil{$^{6}$Department of Physics and Astronomy, University of Leicester, Leicester, LE1 7RH, UK}
\affil{$^{7}$NASA Goddard Space Flight Center, Greenbelt, MD 20771}
\affil{$^{8}$Center for Research and Exploration in Space Science and Technology (CRESST),
Department of Astronomy, University of Maryland College Park, MD 20742-2421}

\begin{abstract}
We present a sample of 27 GRBs with detailed \swift\ light curves supplemented by late time \chandra\ observations. To answer the missing jet-break problem in general, we develop a numerical simulation based model which can be directly fit to the data using Monte Carlo methods. Our numerical model takes into account all the factors that can shape a jet-break: (i) lateral expansion (ii) edge effects and (iii) off-axis effects. Our results provide improved fits to the light curves and constraints on physical parameters. More importantly, our results suggest that off-axis effects are important and must be included in interpretations of GRB jet breaks.

 \end{abstract}

\maxdeadcycles=1000

\section{Introduction}

Since the launch of \swift\ \citep{Gehrels04_Swift_full} in 2004, the nature of jet breaks in gamma-ray burst (GRB) afterglows has become increasingly puzzling. There is strong evidence that the ejecta from the GRB central engine must be jet-like \citep{ZM04}. Thus a collimation correction factor, $f_b=(1-\cos \theta_{jet}) \sim 1/100$, where $\theta_{jet}$ is the jet opening angle with typical values of 5$\degree$-10$\degree$, can be applied to relieve the energy budget problem, so that a typical GRB energy is $E_{\gamma,jet} =(1-\cos\theta_{jet}) $E$_{\gamma,iso} \simeq 0.01 \times 10^{53}= 10^{51} $ erg, where E$_{\gamma,iso}$ is the isotropic equivalent energy release in gamma rays. Such collimated ejecta expand outward relativistically with Lorentz factors $\Gamma$ of several hundred initially. Internally, the ejecta release their energy through internal shocks \citep{1994ApJ...430L..93R,1997ApJ...490...92K,1998MNRAS.296..275D}, or magnetic dissipation processes \citep[e.g, ICMART model;][]{Zhang_Yan_2011} or photospheric dissipation (e.g., Lazzati \& Begelman 2010; Ryde et al. 2010, 2011; Pe'er \& Ryde 2011; Guiriec et al. 2011, 2011, 2015) and produce the prompt $\gamma$-ray emission of GRBs. Externally, the ejecta are further decelerated by an ambient medium (e.g., a constant density interstellar medium, ISM; or a stellar wind environment with density inversely proportional to distance squared) and produce long term broadband afterglows through external shocks \citep[see e.g.,][for a review]{2013NewAR..57..141G}. Due to relativistic beaming, only a portion of the radiation from the ejecta front surface, which is within a cone of half-opening angle $1/\Gamma$, can be observed \citep[][for reviews see Piran 2004, Granot 2007 and van Eerten 2013]{1997ApJ...487L...1R}. An unavoidable consequence of this general picture is that when the ejecta are decelerated to $\Gamma \le 1/\theta_{jet}$, the light curve should steepen because (1) the maximum observable portion of the ejecta (the cone of the whole jet with opening angle $\theta_{jet}$) is now smaller than that which is expected (a cone with half-opening angle $1/\Gamma$ ) and (2) the onset of lateral spreading of the ejecta, predicted to become noticeable in the observer frame around the same time \citep{Rhoads99}, causes the blast wave to decelerate further. Such a ``jet-break" in a GRB light curve is expected to behave achromatically because it only reflects the ejecta geometry, under the assumption that the afterglow emission regions and mechanisms do not change in different spectral regimes (\citealt{Rhoads99, Sari99, Huang2000,Granot02_jet_breaks}; see also reviews by \citealt{Meszaros02, ZM04, Piran04}). The achromaticity was apparently confirmed in the optical and near-IR band in a few cases of pre-\swift\ GRBs \citep{Kulkarni99, Harrison01, Klose04}. 

On the other hand, the number of jet breaks found in \swift\ afterglows is much smaller than expected.
Thanks to the rapid-slew capability of the \swift\ satellite, its X-ray telescope \citep[XRT;][]{Burrows05_XRT} detected nearly 700 X-ray afterglows through 2013 August, typically covering the time ranges from a few minutes to days, weeks, or even months after the GRB trigger times. These data are ideal to test jet-break predictions. Early results, however, suggested that only 
a small fraction \citep[$\sim 12\%$;][]{Racusin09, Liang08} of the \swift\ GRB sample show evidence for canonical jet breaks. This has become known as the ``missing jet break'' problem.

Some promising and natural explanations for the apparent lack of jet breaks in XRT light curves are: (1) jet-breaks exist but are hidden within the data due to uncertainties and observational bias \citep{Curran08,2008ApJ...680..531K,Racusin09} and/or (2) jet-breaks exist but are smoother, less significant in light curves and appear later simply because the GRB ejecta are not pointed directly at us, so that we are observing at an off-axis angle $\theta_{obs}$, where $\theta_{obs} < \theta_{jet}$ \citep{vanEerten10a, vanEerten12}. In either case, much later observations are required to check whether jet breaks occur at or below the \swift/XRT sensitivity limit of a few times $10^{-14}$ erg cm$^{-2}$ s$^{-1}$ (with a typical exposure time of 6 ks). 

These late-time, highly sensitive observations can be carried out by \chandra, which has a limiting flux roughly an order of magnitude lower than the XRT for exposure times of order 60 ks. Over the past several years, we have observed a substantial sample of such late time afterglows using \chandra/ACIS, and the \chandra\ archive includes additional examples. In this paper, by combining those \chandra\ data with \swift/XRT observations and fitting numerical simulations to the resulting light curves, we will address the jet-break problem mentioned above.
 
This paper is organized as follows: we present our data analysis of a sample of GRBs with detailed \swift\ and \chandra\ observations in \S 2. In \S 3, we address the question of whether the smoother/later jet breaks in our sample are due to off-axis observations by directly fitting the numerical model of \citet{vanEerten12} to the observational data with a Monte Carlo Bayesian inference approach. Finally, we summarize our results in \S 4.

\section{Data }

\subsection{Sample Selection}

As of 2013 August, 70 GRBs have been observed by \chandra\footnote{\url{http://cda.harvard.edu/chaser}}, of which 52 triggered the \swift/BAT. 
For the purposes of this study, we want well-sampled \swift\ light curves supplemented by very late-time \chandra\ observations, in order to search for jet breaks at late times that cannot be found in the \swift\ sample alone due to the limited time coverage and/or poor late-time counting statistics of the XRT data.
From this sample of \swift-\chandra\ GRBs we therefore exclude the following bursts: (1) bursts for which the \chandra\ data are not yet available from the archive (GRBs 100814A and 120711A); (2) bursts in which the last data points were not observed by \chandra, so that the light curve is covered adequately by the \swift-only sample (GRBs 051022, 060108, 061021, 060218, 090404, 090407, 100628A, 110312A and 111215A); (3) bursts with poorly-sampled \swift/XRT light curves (GRBs 050412, 050509B, 060505, 100628A, 120624B, 111117A, 100316D, 091117A, 101219A, 101219B and 111020A); (4) bursts affected by possible late time flares or late time shallow decay phases (GRBs 050724, 080913 and 120320A); (5) a burst affected at late times by contamination from a nearby persistent X-ray source \citep[GRB\ 080307;][]{Page09}. Our final sample includes 27 GRBs, which are listed in Table~\ref{tab:main}. 
Five of these bursts (051221A, 060729, 061121, 070125, and 071020) were included in our previous study \citep{Racusin09}, which included the \chandra\ data for the first two of these.
Only two of the GRBs in our sample are short GRBs (defined here as $T_{90} < 2.0$~s). All but three of the bursts in our sample have known redshifts; for those three, we assume $z=2$, which is close to the mean redshift for \swift\ GRBs \citep[see e.g.,][]{2012ApJ...758...46K,2013ApJS..209...20G}.

\begin{table*}

\setlength{\tabcolsep}{2pt}
\caption{The sample of \chandra\ GRBs and the best-fit parameters of the numerical-simulation based model. The uncertainties of the best-fit parameters are calculated at 68\% confidence level.}
\scriptsize
\begin{center}

\begin{tabular}{lccrrrrcccrc}
\hline

GRB & \# & z &$T_{90}$ (s) & $\theta_{jet} $ (deg)& log E$_{53} (erg)$ & log n (cm$^{-3}$)& p & log $ \epsilon_B$ & log $\epsilon_e$ & $\theta_{obs}$ (deg) & $\chi^2/dof$ \\
\hline
\hline
051221A & 1&0.547 & 1.4&$5.08_{-2.49}^{+8.93}$&$1.04_{-1.34}^{+1.39}$&$-1.68_{-1.92}^{+5.02}$&$2.39_{-0.16}^{+0.43}$&$-4.29_{-2.70}^{+1.65}$&$-1.83_{-0.64}^{+1.40}$&$0.10_{-0.10}^{+1.52}$&$40.7/25$\\
060729& 2&0.54 & 115.3&$7.70_{-4.45}^{+5.09}$&$1.58_{-1.35}^{+1.02}$&$-4.14_{-0.86}^{+1.51}$&$2.66_{-0.22}^{+0.05}$&$-2.52_{-1.82}^{+0.87}$&$-1.25_{-0.83}^{+0.79}$&$3.47_{-0.27}^{+3.79}$&$294.2/285$\\
061121& 3&1.314 & 81.3&$4.72_{-0.30}^{+0.53}$&$-0.28_{-0.39}^{+1.22}$&$-2.29_{-0.35}^{+1.23}$&$2.17_{-0.02}^{+0.05}$&$-2.27_{-1.40}^{+0.32}$&$-0.28_{-1.04}^{+0.28}$&$2.84_{-0.05}^{+0.33}$&$175.0/184$\\
070125& 4&1.547\footnote{Fox et al. claimed GRB 070125 is at redshift $z > 1.547$. For simplicty we take z=1.547 for this burst.} & 70&$6.03_{-3.44}^{+5.70}$&$2.39_{-1.85}^{+0.09}$&$2.59_{-0.78}^{+1.81}$&$2.09_{-0.06}^{+0.23}$&$-8.88_{-1.11}^{+2.33}$&$-0.95_{-1.03}^{+0.69}$&$0.15_{-0.15}^{+0.91}$&$44.0/41$\\
071020& 5&2.142 & 4.2&$10.25_{-0.15}^{+4.15}$&$0.52_{-0.76}^{+0.76}$&$2.82_{-0.75}^{+0.75}$&$2.07_{-0.01}^{+0.13}$&$-7.06_{-2.23}^{+0.20}$&$-0.15_{-0.67}^{+0.15}$&$1.60_{-1.02}^{+1.10}$&$81.0/61$\\
080207& 6&2.0858 & 340&$7.48_{-1.69}^{+0.50}$&$1.89_{-1.07}^{+0.76}$&$2.41_{-1.03}^{+0.81}$&$2.27_{-0.05}^{+0.05}$&$-6.02_{-2.71}^{+0.52}$&$-1.81_{-0.71}^{+0.71}$&$4.33_{-0.82}^{+0.73}$&$328.1/75$\\
080319B& 7&0.937 & 50&$5.75_{-3.15}^{+7.82}$&$1.44_{-1.21}^{+1.18}$&$-1.34_{-2.53}^{+3.83}$&$2.22_{-0.19}^{+0.19}$&$-4.63_{-1.49}^{+2.28}$&$-1.55_{-0.83}^{+1.11}$&$0.80_{-0.80}^{+1.39}$&$49.4/55$\\
081007& 8&0.5295 & 10&$25.77_{-0.16}^{+2.60}$&$0.20_{-0.84}^{+0.03}$&$4.61_{-0.55}^{+0.23}$&$2.01_{-0.00}^{+0.01}$&$-6.88_{-0.26}^{+0.56}$&$-0.63_{-0.29}^{+0.47}$&$25.73_{-0.27}^{+0.04}$&$62.7/55$\\
090102& 9&1.547 & 27&$23.92_{-12.08}^{+1.07}$&$0.84_{-0.83}^{+1.66}$&$2.91_{-2.29}^{+0.62}$&$2.43_{-0.07}^{+0.00}$&$-5.86_{-0.63}^{+2.59}$&$-1.05_{-1.40}^{+0.51}$&$19.57_{-3.32}^{+0.59}$&$148.5/137$\\
090113& 10&1.7493 & 9.1&$6.95_{-4.06}^{+7.81}$&$2.16_{-2.13}^{+0.37}$&$1.30_{-3.02}^{+0.61}$&$2.36_{-0.20}^{+0.10}$&$-9.02_{-0.97}^{+2.39}$&$-0.39_{-1.12}^{+0.39}$&$0.02_{-0.02}^{+2.13}$&$14.6/12$\\
090417B& 11&0.345 & $>$ 260&$27.81_{-6.33}^{+0.85}$&$0.83_{-0.23}^{+1.83}$&$3.11_{-1.07}^{+1.07}$&$2.13_{-0.05}^{+0.02}$&$-6.34_{-0.66}^{+3.62}$&$-1.46_{-0.77}^{+1.10}$&$20.90_{-5.41}^{+3.16}$&$182.1/114$\\
090423& 12&8.2 & 10.3&$22.49_{-15.10}^{+0.60}$&$2.59_{-1.87}^{+0.06}$&$0.35_{-0.76}^{+2.47}$&$2.86_{-0.22}^{+0.22}$&$-6.85_{-0.46}^{+3.71}$&$-0.89_{-1.25}^{+0.54}$&$14.58_{-9.86}^{+3.21}$&$23.7/34$\\
091020& 13&1.71 & 34.6&$7.54_{-1.78}^{+14.69}$&$0.64_{-0.42}^{+1.89}$&$0.54_{-2.66}^{+3.32}$&$2.32_{-0.11}^{+0.46}$&$-6.19_{-0.71}^{+3.10}$&$-0.04_{-0.85}^{+0.04}$&$4.18_{-3.03}^{+1.97}$&$91.8/92$\\
091127& 14&0.490 & 7.1&$23.41_{-5.83}^{+3.21}$&$2.20_{-1.44}^{+0.49}$&$-1.53_{-1.76}^{+0.17}$&$2.66_{-0.03}^{+0.03}$&$-5.91_{-0.86}^{+2.65}$&$-0.87_{-0.45}^{+0.68}$&$19.23_{-8.14}^{+4.13}$&$404.5/359$\\
100413A& 15&3.9\footnote{Tentative redshift; Campana et al 2010.} & 191&$7.62_{-2.99}^{+2.15}$&$2.26_{-1.61}^{+0.40}$&$2.93_{-1.69}^{+0.39}$&$2.56_{-0.36}^{+0.03}$&$-6.60_{-0.35}^{+2.27}$&$-1.49_{-0.47}^{+1.12}$&$0.08_{-0.08}^{+1.54}$&$164.4/91$\\
100615A& 16&1.398 & 39&$25.58_{-2.24}^{+2.24}$&$1.29_{-0.74}^{+0.22}$&$4.58_{-0.74}^{+0.25}$&$2.01_{-0.00}^{+0.02}$&$-7.35_{-0.40}^{+0.69}$&$-0.60_{-0.54}^{+0.42}$&$13.65_{-4.43}^{+0.74}$&$64.0/36$\\
100816A& 17&0.8049 & 2.9&$28.16_{-3.73}^{+0.02}$&$1.50_{-1.83}^{+0.71}$&$2.51_{-1.58}^{+1.40}$&$2.01_{-0.00}^{+0.05}$&$-3.38_{-5.22}^{+0.56}$&$-2.82_{-0.58}^{+2.21}$&$0.93_{-0.93}^{+10.92}$&$46.9/27$\\
110402A& 18&2\footnote{Assumed redshift.} & 60.9&$15.02_{-3.68}^{+1.13}$&$-1.27_{-0.09}^{+0.49}$&$4.99_{-0.20}^{+0.01}$&$2.06_{-0.02}^{+0.06}$&$-6.53_{-0.23}^{+0.21}$&$-0.01_{-0.17}^{+0.01}$&$0.56_{-0.56}^{+2.18}$&$26.1/14$\\
110422A& 19&1.77 & 25.9&$16.90_{-1.01}^{+1.32}$&$0.64_{-0.33}^{+0.33}$&$4.85_{-0.65}^{+0.02}$&$2.19_{-0.03}^{+0.02}$&$-6.94_{-0.75}^{+0.75}$&$-0.69_{-0.02}^{+0.54}$&$11.09_{-0.57}^{+0.41}$&$326.9/264$\\
110503A& 20&1.61 & 10&$16.79_{-0.56}^{+2.64}$&$0.48_{-0.52}^{+0.60}$&$4.21_{-0.56}^{+0.56}$&$2.03_{-0.01}^{+0.02}$&$-6.39_{-0.56}^{+0.52}$&$-0.57_{-0.72}^{+0.36}$&$11.85_{-0.71}^{+0.37}$&$436.3/387$\\
110709B& 21&2$^c$ & 846&$26.15_{-3.65}^{+0.06}$&$0.31_{-0.02}^{+0.66}$&$5.00_{-0.35}^{+0.00}$&$2.19_{-0.05}^{+0.05}$&$-7.37_{-1.10}^{+1.10}$&$-0.02_{-0.35}^{+0.02}$&$21.30_{-3.50}^{+0.71}$&$492.5/409$\\
110731A& 22&2.83 & 38.8&$28.66_{-0.72}^{+0.00}$&$0.11_{-0.03}^{+0.59}$&$4.77_{-0.44}^{+0.15}$&$2.10_{-0.01}^{+0.03}$&$-6.18_{-0.24}^{+0.75}$&$-0.10_{-0.61}^{+0.02}$&$24.94_{-0.83}^{+0.37}$&$269.7/224$\\
110915A& 23&2$^c$ & 78.76&$7.53_{-4.93}^{+8.62}$&$0.94_{-0.44}^{+1.67}$&$0.73_{-3.13}^{+0.17}$&$2.46_{-0.00}^{+0.42}$&$-5.96_{-0.45}^{+3.65}$&$-0.26_{-0.77}^{+0.26}$&$3.97_{-2.65}^{+2.05}$&$36.5/39$\\
111008A& 24&4.9898 & 63.46&$3.77_{-1.18}^{+10.07}$&$0.86_{-0.45}^{+1.67}$&$-1.20_{-1.91}^{+2.61}$&$2.35_{-0.18}^{+0.07}$&$-4.58_{-2.40}^{+0.17}$&$-0.05_{-0.57}^{+0.05}$&$2.85_{-0.36}^{+0.36}$&$92.7/100$\\
120804A& 25&1.3 & 0.81&$28.48_{-2.40}^{+0.18}$&$1.73_{-1.06}^{+0.77}$&$3.31_{-1.14}^{+0.72}$&$2.11_{-0.02}^{+0.04}$&$-7.95_{-1.38}^{+1.24}$&$-1.53_{-0.60}^{+1.20}$&$27.75_{-0.39}^{+0.17}$&$70.1/31$\\
121027A& 26&1.773 & 62.6&$19.10_{-0.02}^{+7.92}$&$1.77_{-0.98}^{+0.74}$&$2.70_{-0.40}^{+1.69}$&$2.16_{-0.05}^{+0.13}$&$-6.49_{-2.41}^{+0.54}$&$-1.62_{-0.23}^{+1.31}$&$11.65_{-1.77}^{+1.77}$&$80.1/72$\\
130420A& 27&1.297 & 124&$25.87_{-1.09}^{+1.09}$&$0.27_{-0.48}^{+0.29}$&$4.74_{-0.40}^{+0.16}$&$2.02_{-0.01}^{+0.02}$&$-7.04_{-1.07}^{+0.16}$&$-0.38_{-0.26}^{+0.28}$&$25.77_{-4.29}^{+0.09}$&$105.2/82$\\

\hline
\end{tabular}

 \end{center}
\label{tab:main}
\end{table*}

\subsection{Data Reduction}

There are two sets of data for each GRB in our sample: \swift/XRT data and \chandra/ACIS data. 
To account for the spectral evolution which is sometimes observed in the XRT light curves, we used the 
Swift/XRT team's standard ``Burst Analyzer"\footnote{\url{http://www.swift.ac.uk/burst\_analyser/}} results for each burst. 
In those analyses, the count-to-flux conversion factor\footnote{The uncertainties associated with the conversion of counts to flux, however, were not included in this analysis.} is time-dependent and accounts for spectral evolution. A  customized code was written 
to generate the X-ray flux density at 1 keV. Interested readers are referred to \cite{2007A&A...469..379E,Evans09} for more technical details.

All the \chandra\ observations were obtained with the ACIS instrument in Very Faint (VF) mode. We first obtained the count rate by the following steps: (1) we processed the \chandra\ data using the \chandra\ Interactive Analysis of Observations (CIAO; version 4.5) software and the calibration database (CALDB; version 4.5.6) downloaded from the official \chandra\ 
website\footnote{\url{http://cxc.harvard.edu/ciao/index.html}}, 
reprocessing all the data using the automated script {\tt chandra\_repro} to ensure that the latest calibration files have been applied. 
(2) Next we extracted the event files from the source and background regions, respectively, using {\tt dmextract}. 
Depending on the brightness of the \chandra\ afterglow source, we selected a circular source extraction region of radius 0.8-1.2 arc seconds, and used a source-free background region of 20 arc seconds radius located far from the source in the CCD image. (3) We then calculated the net count rate in the energy range 0.2-8 keV by dividing the background-subtracted photon counts by the exposure time. Source and background spectra were extracted using {\tt specextract} from the source and background event files, respectively. The 1-$\sigma$ uncertainties of the \chandra\ observations are estimated using \cite{Kraft91}, which gives the method to calculate confidence limits for experiments with low numbers of counts. Upper limits in counts are first calculated from the source and background counts using Kraft et al. 1991, then converted to flux units.

The conversion of the \chandra\ count rate to flux density at 1 keV depends in principle on the spectral shape in the \chandra\ band. Since late-time \chandra\ photon counts are in most cases very low, the \chandra\ data generally can not constrain this spectral shape. In order to obtain the count-to-flux factor, we therefore fit the \chandra\ data to absorbed power laws with $\Gamma_X$ and N$_H$ fixed to the nearest XRT values in time that are available in the Burst Analyzer results mentioned above. Then we convert the \chandra\ count rate in 0.2-8 keV to flux density at 1 keV. We finally combine the Swift/XRT and \chandra/ACIS observations and present the unabsorbed light curve of each burst at 1 keV. As shown by Tsujimoto et al. 2011, both instruments agree to within about 10\%, and ignoring inter-calibration uncertainties will therefore not affect our results.

{ For the purpose of this study, we only focus on the late-time afterglow-dominated X-ray data that do not overlap with the prompt emission, steep decay phase \citep{2005Natur.436..985T,2007ApJ...666.1002Z,2009ApJ...690L..10Z}, shallow decay phase \citep{Nousek06,Zhang06,2007ApJ...670..565L} or X-ray flares \citep{2005Sci...309.1833B,Falcone:2008wz,2010MNRAS.406.2113C,2010MNRAS.406.2149M,2011MNRAS.410.1064M}. We thus selected the fitting time interval by the following steps: (1) measure $t_{\rm burst}$ according to \cite{2013arXiv1310.2540Z}, where $t_{\rm burst}$ is defined as the time during which the observed ($\gamma$-ray and X-ray) 
emission of a GRB is dominated by emission from a relativistic jet via an internal
dissipation process (e.g. internal shocks or magnetic dissipation) and 
not by the afterglow emission from the external shock. (2) If the X-ray light curve after  $t_{\rm burst}$ can be fit with a single power-law, we use the whole time interval after $t_{\rm burst}$ for our fits.  (3) Otherwise, if the X-ray light curve after $t_{\rm burst}$ is fitted by two or more power-law segments we took the {\it last two} power law segments as our fitting interval for the remainder of this paper. The results of our fits are presented in the on-line material (see the fit for GRB 111008A shown in Figure 1, for an example). In each plot \chandra\ data points are plotted in red while \swift/XRT data points are plotted in green and gray, with gray used to show points that were excluded from our fits. The \chandra\ data points have expanded both the temporal and flux coverage by up to an order of magnitude.

\section{A Physical Model Based on Numerical Simulations}

 GRB afterglow theory indicates that there are two distinct effects that contribute to jet breaks: (i) lateral spreading of the collimated outflow, which reduces the energy per unit solid angle in the jet by increasing its solid angle; and (ii) the effects of the edge of the jet becoming visible as the ejecta decelerate and the relativistic Doppler beaming decreases. In addition, the angle between the jet axis and the observer, $\theta_{obs}$, has important effects on the timing, sharpness, and slope change of a jet break.
Ideally, any successful model should take into account all three factors to explain the observed light curves. Numerical simulations (Kumar \& Granot 2003, Meliani et al. 2007, Zhang \& MacFadyen 2009, van Eerten et al. 2011, Wygoda et al. 2011, De Colle et al 2012) show that, in practice, jets spread out sideways closer to logarithmically than exponentially (as initially predicted by Rhoads 1999), due to the quick transition into the trans-relativistic regime for realistic opening angles. Off-axis light curves from simulations (van Eerten et al. 2010, van Eerten \& MacFadyen 2012a, van Eerten \& MacFadyen 2012b) indicate that the jet break is shaped jointly by jet spreading and observer angle, confirming the importance of jet orientation relative to the observer for the shape of the light curve. In this section, we use the results of numerical simulations to directly fit the observational data and constrain the jet properties.

\subsection{The Numerical Simulations}

The relativistic hydrodynamics (RHD) simulations we use have been discussed extensively in 
\citet{ZhangW_MacFadyen_2009}, \citet{vanEerten10a}, \citet{vanEerten12}, and 
\citet{vanEerten_MacFadyen_2013}. 
We will therefore only summarize some of the important points which are relevant to this work: 

- The two-dimensional hydrodynamics (RHD) simulations: As initial condition at a time prior to causal contact in the angular direction of the flow, the simulations take a top-hat Blandford-McKee profile (Blandford \& McKee, 1976), truncated at a given opening angle.  The radiation and the dynamics of the blast wave are assumed to be separated, which generally remains valid for GRBs since the efficiency and feedback effects are typically small in GRB jets. The fraction of energy carried by magnetic fields at the front of the blast wave is also assumed to be small. The RHD simulations are performed with an adaptive-mesh refinement algorithm that allows grid resolution to vary across the computational domain so that computational resources can be concentrated where they are most needed
\citep{ZhangW_MacFadyen_2006}. 

- Scale-invariant initial conditions: Fortunately, the computational cost can be dramatically reduced when taking advantage of the scale invariance of the jet evolution \citep{vanEerten12,vanEerten_MacFadyen_2012a}.
For an explosion with isotropic equivalent energy E$_{iso}$ occurring in a homogeneous medium with density $\rho_0$, the fluid state at a distance $r$, angle $\theta_{obs}$, and source frame time $t_e$ can be simply determined by a small number of independent dimensionless combinations of the variables: $A\equiv r/ct_e$, $B\equiv E_{iso}t_e^2/\rho_0r^5$ and $\theta_{jet}$. Thus the critical parameter that determines the ``dimensionless" jet properties is $\theta_{jet}$. 
For a given $\theta_{jet}$, simulations for any arbitrary values of 
$\{r^\prime, t^\prime_e, {\rm and}~(E_{iso}/n_0)^\prime\}$,
where $n_0\approx\rho_0/m_p$ for proton mass $m_p$,
can be derived from a single simulation that calculates the jet structure as a function of $t_e$ for parameters $\{A^\prime~\rm{and}~B^\prime\}$. In practice, we have run 19 simulations with $\theta_{jet}$ ranging from 0.045 to 0.5 \citep[see Table~\ref{tab:main} in][]{vanEerten12}.

- Table-based radiation calculations: 
The scaled 2D simulation can then be used to calculate the radiation transfer along a line of sight at angle $\theta_{obs}$. 
Radiation transfer is calculated following \citet{1998ApJ...497L..17S} and \citet[][see Appendix A]{vanEerten12}. We only consider the ISM case throughout this paper. 
The radiative transfer calculation is computationally expensive, so these calculations are performed on a grid of parameter values and stored as tables that can be used later to fit the model to observational data.

The characteristic synchrotron spectral shape is determined by the peak flux, $F_{peak}$, the synchrotron break frequency, $\nu_{m}$, and the cooling break frequency,  $\nu_c$. These can be expressed in terms of scale invariant functions $\mathfrak{F}_{peak}$, $\mathfrak{f}_{m}$ and $ \mathfrak{f}_{c}$ respectively, which in turn are functions of $\mathcal{T}_{obs}$, $\theta_{jet}$ and $\theta_{obs}$. Here $\mathcal{T}_{obs}$ is a scaled time that combines $E_{iso}/n_0$ with the observer time (accounting for redshift explicitly and not including it in $\mathcal{T}_{obs}$) as follows:

\begin{equation}
\mathcal{T}_{obs}\equiv (\frac{n}{E_{53}})^\frac{1}{3}\frac{t_{obs}}{1+z} \ .
\label{eq:6}
\end{equation}
Here $z$ is the redshift, and the energy and density scale factors, $E_{53}$ and $n$, are defined as:
\begin{equation}
E_{53} \equiv \frac{E_{iso}}{10^{53} {\rm erg}}
\label{eq:7}
\end{equation}
and 
\begin{equation}
n\equiv \frac{n_0}{1\text{\ } {\rm cm}^{-3}} \ .
\label{eq:8}
\end{equation}

Following van Eerten \& MacFadyen 2013, we have tabulated $\mathfrak{F}_{peak}$, $\mathfrak{f}_{m}$ and $\mathfrak{f}_{c}$, which are derived directly from the numerical simulations.

Then the {\textbf{observed}} spectral shape can be calculated as:
\begin{equation}
 F_{peak} = \frac{(1+z)}{d^2_{28}} \frac{p-1}{3p-1} \epsilon_B^{1/2} E_{53} n^{1/2} \mathfrak{F}_{peak} (\mathcal{T}_{obs} ; \theta_{jet}, \theta_{obs})
 \label{eq:9}
\end{equation}
\begin{equation}
 \nu_{m} = (1+z)^{-1} \left( \frac{p-2}{p-1} \right)^2 \epsilon_e^2 \epsilon_B^{1/2} n^{1/2} \mathfrak{f}_{m} (\mathcal{T}_{obs} ; \theta_{jet}, \theta_{obs})
 \label{eq:10}
\end{equation}
\begin{equation}
 \displaystyle \nu_{c} = (1+z)^{-1} \epsilon_B^{-3/2} E_{53}^{-2/3} n^{-5/6} \mathfrak{f}_{c} (\mathcal{T}_{obs} ; \theta_{jet}, \theta_{obs})
 \label{eq:11}
\end{equation}
where $\epsilon_e$ and $\epsilon_B$ are electron energy density fraction and magnetic field energy density fraction respectively, $p$ is the power-law index of the emitting electrons, and $d_{28}$ is the luminosity distance in units of $10^{28}$~cm \citep{vanEerten_MacFadyen_2013}.\footnote{In their original form, Eqs.~\ref{eq:9} and \ref{eq:10} had an additional multiplicative variable, $\xi_N$, which denotes the fraction of the downstream electron number density that participates in the shock-acceleration process. Following common practice for afterglow analysis, we assume $\xi_N = 1$ throughout this work, and have therefore omitted this variable from the equations.}

Depending on the relation between $\nu_m$ and $\nu_c$, we can further calculate the observed flux density $F_\nu$ at any observed frequency $\nu$ following either the slow-cooling ($\nu_c > \nu_m $) or fast-cooling 
($\nu_c < \nu_m$) spectral regimes \citep{Sari98}:

\begin{equation}
F_{SLOW}(\nu)=
\begin{cases}
F_{peak}(\frac{\nu}{\nu_m})^{\frac{1}{3}}; \ \textrm{$ \nu<\nu_m<\nu_c$} \\
F_{peak}(\frac{\nu}{\nu_m})^{\frac{1-p}{2}}; \ \textrm{$ \nu_m<\nu<\nu_c$} \\
F_{peak}(\frac{\nu_c}{\nu_m})^{\frac{1-p}{2}} (\frac{\nu}{\nu_c})^{-\frac{p}{2}}; \ \textrm{$ \nu_m<\nu_c<\nu$} 
\end{cases}
\label{eq:12}
\end{equation}

\begin{equation}
F_{FAST}(\nu)=
\begin{cases}
F_{peak}(\frac{\nu}{\nu_c})^{\frac{1}{3}}; \ \textrm{$ \nu<\nu_c<\nu_m$} \\
F_{peak}(\frac{\nu}{\nu_c})^{-\frac{1}{2}}; \ \textrm{$ \nu_c<\nu<\nu_m$} \\
F_{peak}(\frac{\nu_m}{\nu_c})^{-\frac{1}{2}} (\frac{\nu}{\nu_m})^{-\frac{p}{2}}; \ \textrm{$ \nu_c<\nu_m<\nu$} 
\end{cases}
\label{eq:13}
\end{equation}

In summary, in addition to the three hydrodynamics model parameters, the radiation calculations introduce 4 new parameters ($\{\theta_{obs}, p, \epsilon_e, \epsilon_B\}$): the off-axis observation angle and the three synchrotron radiation microphysical parameters. By using the above procedure, we can fully calculate the observed flux density $F_\nu$ at any observed frequency $\nu$ and observer time $\mathcal{T}_{obs}$ based on the numerical tables $\mathfrak{F}_{peak}$, $\mathfrak{f}_m$, and $\mathfrak{f}_c$ . Effectively, we define this calculation of the computational model as:
\begin{equation}
F_\nu(\mathcal{T}_{obs})=M(\nu,\mathcal{T}_{obs}; \{\theta_{jet}, E_{53}, n, p,\epsilon_e,\epsilon_B, \theta_{obs}\} ) \ .
\label{equ:model}
\end{equation}

The three numerical tables, $\mathfrak{F}_{peak}$, $\mathfrak{f}_m$, and $\mathfrak{f}_c$, were calculated on grids of 100 values in the parameters $\mathcal{T}_{obs}$, $\theta_{jet}$, and $\theta_{obs}/\theta_{jet}$ (Eqs.~3, 4, and 5). 
The tables use parameter ranges of $86.4 \le \mathcal{T}_{obs} \le 8.64 \times 10^8$, $0.045 \le \theta_{jet} \le 0.50$, and $0.0 \le \theta_{obs}/\theta_{jet} \le 1.0$, and are scaled and interpolated from the original 19 detailed simulations.
These tables allow us to quickly fit Eq.~\ref{equ:model} to the observational data. 

\subsection{Limitations of the Numerical Model Tables}

The current set of simulation-templates have all been calculated assuming a homogeneous circumburst medium. An obvious alternative would be using a stellar-wind type profile with radially decreasing density, as expected from massive star progenitors. Nevertheless, studies have shown that a large number of GRBs afterglows are described best using an interstellar medium (ISM) type density (see e.g. Panaitescu \& Kumar 2001, 2002; Racusin et al. 2009, Curran et al. 2011; Cenko et al. 2011; Schulze et al. 2011). A larger set of templates including wind profiles is in preparation (wind profiles require specialized numerical techniques to fully resolve, such as described in Van Eerten \& MacFadyen 2013).

Another limitation in the tables is that they have been computed using a hybrid approach to electron cooling \citep{ZhangW_MacFadyen_2009}, where a single global cooling time (equated to burst duration) is taken to apply to the entire fluid profile \citep{1998ApJ...497L..17S}. The core afterglow fluid dynamical properties remain unchanged, such as jump conditions at the shock front and spreading behavior. As a result, this approach generates all the correct scalings and temporal evolution, compared to a local approach to electron cooling time. It differs however by a dimensionless integration factor provided by the downstream radial cooling time profile (which is flat in the global cooling case), introducing an essentially constant relative shift in cooling break $\nu_c$ \citep{vanEerten10a}. As shown for example by \cite{2014MNRAS.438..752G}, this can have a significant effect on broadband fit results that constrain model parameters $E_{iso}$, $n_0$, $\epsilon_B$ and $\epsilon_e$. In the current work we limit ourselves to a single band (X-rays) and do not constrain these parameters typically beyond order of magnitude estimates at most, at which order the impact of the cooling break shift is negligible. The other model parameters, $p$, $\theta_0$ and $\theta_{obs}$ are more sensitive by far to light curve curvature rather than flux level and are marginally or not at all affected by the global vs. local cooling time issue \citep[this was also tested explicitly and confirmed in the context of][]{2015ApJ...799....3R}. Nevertheless, like the fundamental model degeneracy introduced by $\xi_N$\citep{2005ApJ...627..861E}, these are issues to be kept in mind when interpreting fit results. It should be noted that these issues occur as well when interpreting afterglow curve fits in terms of analytical models of blast waves and jet breaks, where issues like $\xi_N$, radial fluid profile and observer angle are typically ignored completely.

Equations~\ref{eq:9} -- \ref{eq:13} were derived under the assumption that $p \ge 2$, which is carried as an implicit assumption in our results. Lower values of the electron energy index require a high energy cutoff to avoid divergence of the integrated electron energy; \citet{Dai_Cheng_2001} and \citet{Racusin09} consider the case of $1 < p < 2$, which has been assumed for some bursts, but since this introduces at least one additional parameter for the high energy cutoff we do not consider that case here.

The numerical model tables also impose cutoffs in the values of the model parameters that we are able to fit. 
Of particular interest in this regard is the jet opening angle. Because of the requirement of starting prior to causal contact across jet angles, very narrow jets have to be started at extremely early times and corresponding high Lorentz factors, which becomes computationally prohibitive due to the resolution needed in order to resolve the Lorentz-contracted radial shell width; as a result we limited our simulated jets to opening angles $\theta_{jet} \ge 0.045$~radians. Since previous studies found typical jet angles of 0.05 -- 0.1 radians (see e.g., Harrison et al. 1999, Stanek et al. 1999, Frail et al. 2001), we also limited our model grid to opening angles $\theta_{jet} \le 0.5$~radians.

\subsection{A Monte Carlo Fit} 

Our next step is to fit the numerical simulation model (eq. \ref{equ:model}) to the observational data. As discussed in \S 2.2, all the observed light curves have been{ converted} to{ flux densities at} 1 keV. Thus our model to fit the data is actually (from eq. \ref{equ:model} by setting $h\nu= 1~\rm{keV} $):
 \begin{equation}
F_{1\rm{keV}}(\mathcal{T}_{obs})=M(\mathcal{T}_{obs}, \mathcal{P} )
\label{equ:fit}
\end{equation}
where $\mathcal{P}$ stands for the parameter set of $\{\theta_{jet}, E_{53}, n, p,\epsilon_e,\epsilon_B, \theta_{obs}\} $.

Due to the complexity of the problem, the insensitivity of the light curve to some input parameters, degeneracy of some parameters and possible multi-modality of the likelihood in the parameter space, it would not be appropriate to apply the commonly used least $\chi^2$ fitting algorithms to constrain the model parameters. Instead we used the widely used {\it MULTINEST} \citep{2008MNRAS.384..449F,2009MNRAS.398.1601F,2013arXiv1306.2144F}, a Bayesian inference tool that can explore the complete parameter space efficiently. By generating a full posterior probability distribution
function (PDF) of all the parameters using {\it MULTINEST}, we are able to evaluate the best-fit parameters and their uncertainties. If
the posterior PDF is multi-modal, we choose the region that contains the highest likelihood values (i.e, smallest $\chi^2$ values). 
We cross checked the results by using our own MC fit codes, which are based on the Affine Invariant Markov chain algorithm \citep{Goodman_Weare_2010} with parallel tempering enabled \citep[see, e.g.,][]{2005PCCP....7.3910E} and found they are consistent with each other. {\it MULTINEST}  was chosen based on its relative computational performance. For the details of the approach using parallel tempering MCMC, see \cite{2015ApJ...799....3R}, who analyzed a larger Swift-only sample of 
XRT light curves. When applied to the same data, the two independently 
developed statistics codes, MCMC with parallel tempering from \cite{2015ApJ...799....3R} and multimodal nested sampling from the current work, yield 
consistent results, confirming both the conceptual validity and the 
practical implementations of our methods.

We note that there are indeed correlations and some degree of degeneracy between some parameters in our problem. An obvious example is that $E_{53}$, $n$, $\epsilon_e$, and $\epsilon_B$ are highly correlated with each other (e.g., see Figure~3 in on-line materials). Because of these correlations and degeneracy, it is not possible to constrain all the parameters in Eq.~\ref{equ:fit} using a single energy band (i.e, X-ray) data. 
{ Indeed, these parameters are often poorly constrained and highly correlated in our fits, but that does not prevent us from fitting the jet opening angle, jet viewing angle, and electron energy index for most afterglows.}

The allowed ranges of the fitting parameters are listed in Table \ref{tab:range}. We also require $\epsilon_B < \epsilon_e $ for each burst,{ which is expected from both theoretical \citep[e.g.,][]{2006ApJ...651L...9M} and numerical calculations \citep[e.g.,][]{2009ApJ...707L..92S}}.

\begin{table}

\setlength{\tabcolsep}{2pt}

\caption{Allowed Range of the Numerical Fitting parameters}
\begin{center}

\begin{tabular}{l|l}
\hline
Parameter & Range \\
\hline
$\theta_{jet}$ (rad) & [0.045, 0.5] \\
$E_{53}$ (erg) & [$10^{-10}$, $10^{3}$] \\
 $n$ (cm$^{-3}$) & [$10^{-5}$, $10^{5}$]\\
p & [2, 4]\\
$\epsilon_B$ & [$10^{-10}$, 1] \\
$\epsilon_e$ & [$10^{-10}$, 1] \\
$\theta_{obs}/\theta_{jet}$ & [0, 1] \\
\hline

\hline
\end{tabular}

\end{center}

 \label{tab:range}
\end{table}

\subsection{Afterglow Fits}

The fitting results are presented in Table~\ref{tab:main}. An example of the distributions of parameter values is shown for GRB 111008A in Fig \ref{fig-111008A} (the complete fit results are included in the on-line material). Here we plot the converged {\it MULTINEST} results for all parameters, marginalized over all except one or two in all possible combinations. The uncertainties on the parameters shown in Table~\ref{tab:main}, which are the 68\% uncertainties of the local mode region that includes the best-fit parameters, were obtained from the {\it MULTINEST} outputs.

\begin{figure*}
 \centering
 \includegraphics[width=0.95\textwidth]{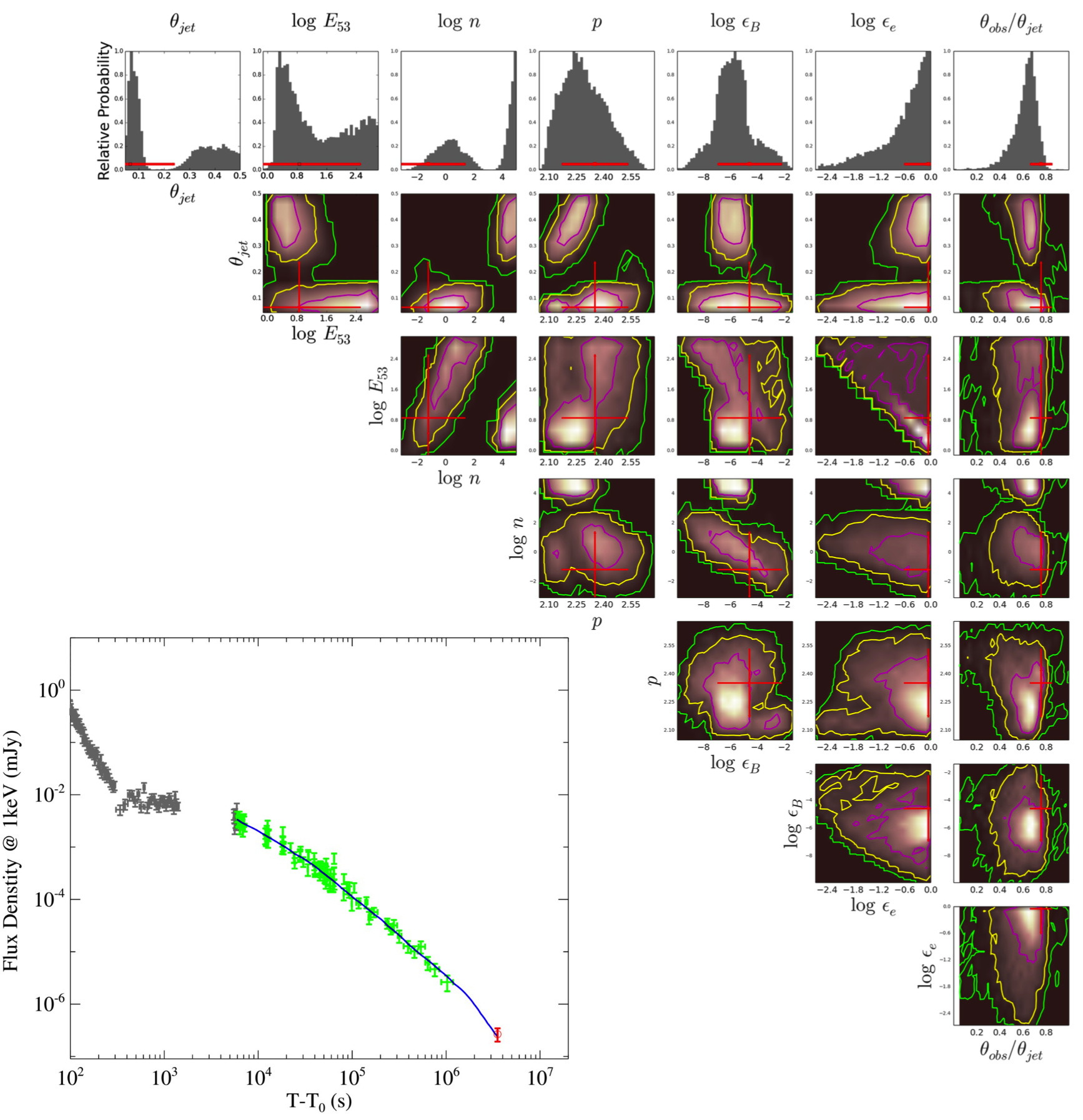}
 \caption{
{\scriptsize Fit of our simulation-based model to the data for GRB 111008A. In the bottom-left corner plot, gray and green points are {\it Swift}/XRT data points. Red open circles are {\it Chandra} observed data points. Blue solid lines represent the modeled light curves, which are fit to only the green and red data points. The top row shows histograms for individual models parameters. The remaining panels show contour plots for pairs of model parameters of the relative probability marginalized over the remaining model parameters, determined from the converged MC chains in our numerical simulation-based model. Red crosses indicate the best-fit values listed in Table~\ref{tab:main}. Length of solid lines represents the uncertainties of the parameters listed in Table~\ref{tab:main}. The best-fit points and uncertainties are derived from the best simultaneous fit to all of the model parameters, and therefore do not necessarily correspond to the peak probabilities in the 1-D histograms or 2-D contour plots of the relative marginalized probabilities.}}
\label{fig-111008A}
\end{figure*}

\bigskip
We consider each light curve briefly here.

\paragraph{GRB 051221A} This is one of only two short GRBs in our sample. Soderberg et al. 2006 gives a jet angle of 7\degree. The later \chandra\ data clearly indicate the presence of a jet break \citep[][]{Burrows06_GRB051221A}.  The distributions of parameter values from the fits, as well as contours of our fitting statistic for pairs of model parameters are show in the plot in the online material. We note that both angles have fairly large uncertainties but they generally suggest a off-axis jet. The posterior distributions are multi-modal and multiple solutions exist that give the equally good fits to the data. In order to give a comparable ``best-fit" result as traditional fitting techniques do, we pick (and thereafter for other GRBs in this paper as well) ``best-fit" values corresponding to the highest likelihood in the final converged MC chains, which is, in other words, the global maximum peak of the likelihood as function of the 7 parameters. The uncertainties of those best-fit values  are estimated based on the size of local modes (``likelihood islands") in the parameter space. Thus such uncertainties are typically underestimated if the solution is multimodal. In this approach, the numerical fits give a narrow, but poorly-constrained jet,{ with $\theta_{jet} \sim 5.1 \degree$ and an off-axis angle of $ 0.1\degree$ (Table~\ref{tab:main})} for this burst.
 
\paragraph{GRB 060729} The is the longest GRB afterglow in our sample, and the extremely late-time \chandra\ observations \citep{Grupe10} should provide excellent leverage to determine the jet parameters. The numerical model finds a solution with best-fit parameters of $\theta_{jet} \sim 7.7\degree$ and $\theta_{obs} \sim 3\degree$ (Table~\ref{tab:main}). A large ratio of $\theta_{obs}/\theta_{jet}$ $\sim$ 0.5 suggests that the jet is likely off-axis and results in a  late-time break in the slope. We note that multi-modal features in the posterior distribution are clearly seen. So an acceptable good fit with on-axis jet may also work for the data. The parameters mentioned above are the best-fit parameter with maximum likelihood as discussed in \S 3. Further study including multi-wavelength data may help reduce the multi-modality. 

\paragraph{GRB 061121} The \chandra\ data point falls below the extrapolation of the \swift\ light curve. The jet break  is smoothed and moderated in the numerical model with an opening angle of $4.7 \degree$, and an off-axis angle of about $3.5 \degree$.} The histograms for $\theta_{jet}$, $p$, and $\theta_{obs}/\theta_{jet}$ are quite narrow .{ The other parameters are poorly constrained with flat-shaped histograms.}

\paragraph{GRB 070125} The numerical fit is poorly constrained for both $\theta_{jet}$ and $\theta_{obs}$, which span their allowed range in parameter space.

\paragraph{GRB 071020} A late-time deep \chandra\ upper limit leads the numerical fit to a smooth break. The jet has an opening angle of $10.3 \degree$, and an off-axis angle of about $1.6 \degree$.

\paragraph{GRB 080207} The late-time \chandra\ data point is consistent with the \swift\ light curve. The numerical fit favors a jet angle of $7.5 \degree$ and a large viewing angle. The electron index, $p$, is well-constrained for this burst.

\paragraph{GRB 080319B} \citet{Racusin08} found that this ``naked-eye'' burst required a two-component jet to fit the detailed multi-wavelength data set. Here we fit only the late-time X-ray data to a single jet model. The numerical model obtains a jet angle of ${5.8\degree}_{-3.2}^{+7.8}$ \citep[c.f. $4.0\degree$ found by][]{Racusin08}. We also obtain a high density and an electron index of about 2.2. The posterior distribution is clearly multi-modal.

\paragraph{GRB 081007}. The numerical fit gives a large off-axis jet with poorly-constrained jet opening angle $\sim 25.8 \degree$ and $\theta_{obs} \sim 25.7 \degree$.

\paragraph{GRB 090102} The two late-time \chandra\ data points increase the light curve duration by an order of magnitude. These late-time \chandra\ data points are consistent with the \swift\ light curve, leading to a late-time decay slope of $\sim 1.5$ and a lower limit on the opening angle. The numerical model parameter distributions show a broad distribution of jet angles and favor large viewing angles. Only the electron distribution index, $p$, is well constrained at $\sim$ 2.4.

\paragraph{GRB 090113} This is another light curve consistent with a simple power law. The jet opening angle and observing angle are poorly constrained.

\paragraph{GRB 090417B} This afterglow can be fitted with a broad off-axis jet. Our single best fit result indicates a large angle, but when taking into account the parameter space as a whole, the largest probability is assigned to angles clustered around a smaller value. The electron index is about 2.1.

\paragraph{GRB 090423} The \chandra\ data extend the \swift\ light curve by nearly an order of magnitude in time. The jet opening angle and observing angle are poorly constrained.

\paragraph{GRB 091020} The \chandra\ data point falls below the extrapolation of the \swift\ light curve, resulting in a curved numerical light curve. The parameter distributions { weakly} favor a jet angle of about $7.5\degree$ with large uncertainties and a poorly constrained off-axis angle. The other model parameters are unconstrained, with distribution functions covering an order of magnitude.

\paragraph{GRB 091127} The light curve can be fitted by a broken power law with an early break and is achromatic with optical observations \citep{2012ApJ...761...50T}, but the slight curvature of the numerical model improves the fit) and favors a broad, off-axis jet. The rest of the jet parameters (other than $p \approx 2.7$) are very poorly constrained. 

\paragraph{GRB 100413A} The jet parameters are constrained to the region of $\theta_{jet}$ $\le$ 11.5 \degree (0.2 radian).   The $p$ distribution shows a bimodal feature between 2.2 and 2.6. The fit favors an off-axis jet.

\paragraph{GRB 100615A}  The numerical solution favors a jet with large opening and off-axis angles.

\paragraph{GRB 100816A} The numerical solution favors a wide, on-axis jet. All the jet parameters, except for $p$ ($\sim $ 2.0) are distributed with quite large uncertainties. The posterior distribution for jet orientation is bi-modal.

\paragraph{GRB 110402A} The final data point is a deep \chandra\ upper limit that falls slightly below the extrapolation of the \swift\ data points. The numerical fit accounts for this upper limit through curvature of the light curve into the trans-relativistic regime. The distribution of $\theta_{jet}$ is somewhat bimodal although a solution of  $\theta_{jet}$ = $15 \degree$ is favored. The observer's angle, $\theta_{obs}$,  is unconstrained but corresponds to an on-axis solution for $\theta_{jet}$ = $15 \degree$.

\paragraph{GRB 110422A} The numerical fit obtains a large jet opening angle of $16.9\degree$ and favors an off-axis view angle. $p$ is constrained to about $2.1_{-0.02}^{+0.06}$. Other parameters (except for $p$) are poorly constrained to about an order of magnitude.

\paragraph{GRB 110503A} { The numerical result gives a well-constrained large jet-angle of $16.8\degree$ and well-constrained viewing angle of $11.9\degree$. The electron index is about 2.0.

\paragraph{GRB 110709B} This burst has a long, well-measured \swift\ light curve with a single late \chandra\ point at $t\sim 10^7$ s. A large jet angle with large observer angle is favored.

\paragraph{GRB 110731A} This burst was also detected by Fermi/LAT\citep{2013ApJ...763...71A}. Three late-time \chandra\ observations provide evidence for a jet break at about 7.5~Ms after the burst, and $p$ $\sim$ 2.1. A large jet angle and large observer angles are favored.

\paragraph{GRB 110915A}  The numerical fit produces a slightly curved light curve, but not sufficiently to constrain any of the model parameters, leaving viewing angle and opening angle essentially unknown.

\paragraph{GRB 111008A} This burst has the highest redshift in the sample. The \chandra\ flux density lies sufficiently below the extrapolation of the Swift light curve to indicate the presence of a jet break.. The numerical fit determines a jet angle of about $3.8\degree$ with an off-axis angle of $2.9\degree$.

\paragraph{GRB 120804A} One of only two short GRBs in our sample, this afterglow has two late \chandra\ observations that are consistent with the \swift\ light curve. The numerical fit favors a large jet angle and a viewing angle close to the jet edge.

\paragraph{GRB 121027A} This burst has a dramatic X-ray flare lasting from $\sim 1-30$~ks. The numerical fit favors a large jet angle of $\sim 19 \degree$ viewed off-axis.

\paragraph{GRB 130420A} The numerical fits suggest a large jet angle of 26$\degree$ viewed off-axis.

\subsection{Summary of Results and Discussion}

Previous studies have typically fitted broken power laws to the afterglow light curves and used the break time to determine an estimate for the jet angle. Some studies \citep[e.g,][]{2001ApJ...560L..49P,2002ApJ...571..779P,2003ApJ...597..459Y,2013ApJ...776..119L,2014ApJ...781...37P} have used fits of analytic afterglow models to estimate other parameters as well. The advantage of our direct fit of numerical models is that we can account for non power-law light curve shapes, and that the physical model parameters can be constrained directly using as much information as possible from the data. Furthermore, the Monte Carlo method enables us to find confidence intervals for the model parameters (Table~\ref{tab:main}) by using the distributions found from the fitting procedure.

We summarize the key points as follows:

\begin{figure}[ht!]
\begin{center}
\subfigure[ ]{
\includegraphics[scale=0.9]{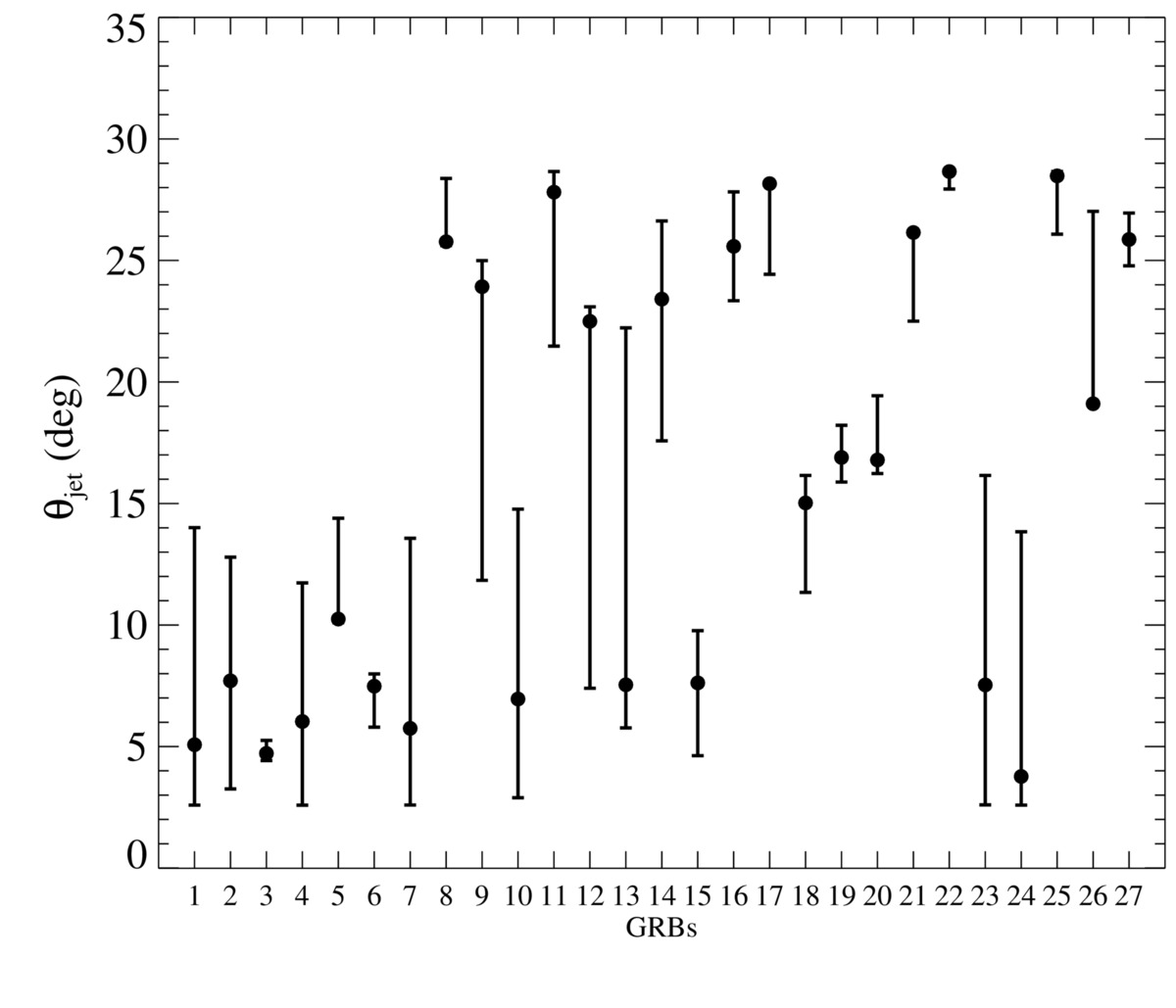}
 }
\subfigure[ ]{
\includegraphics[scale=0.9]{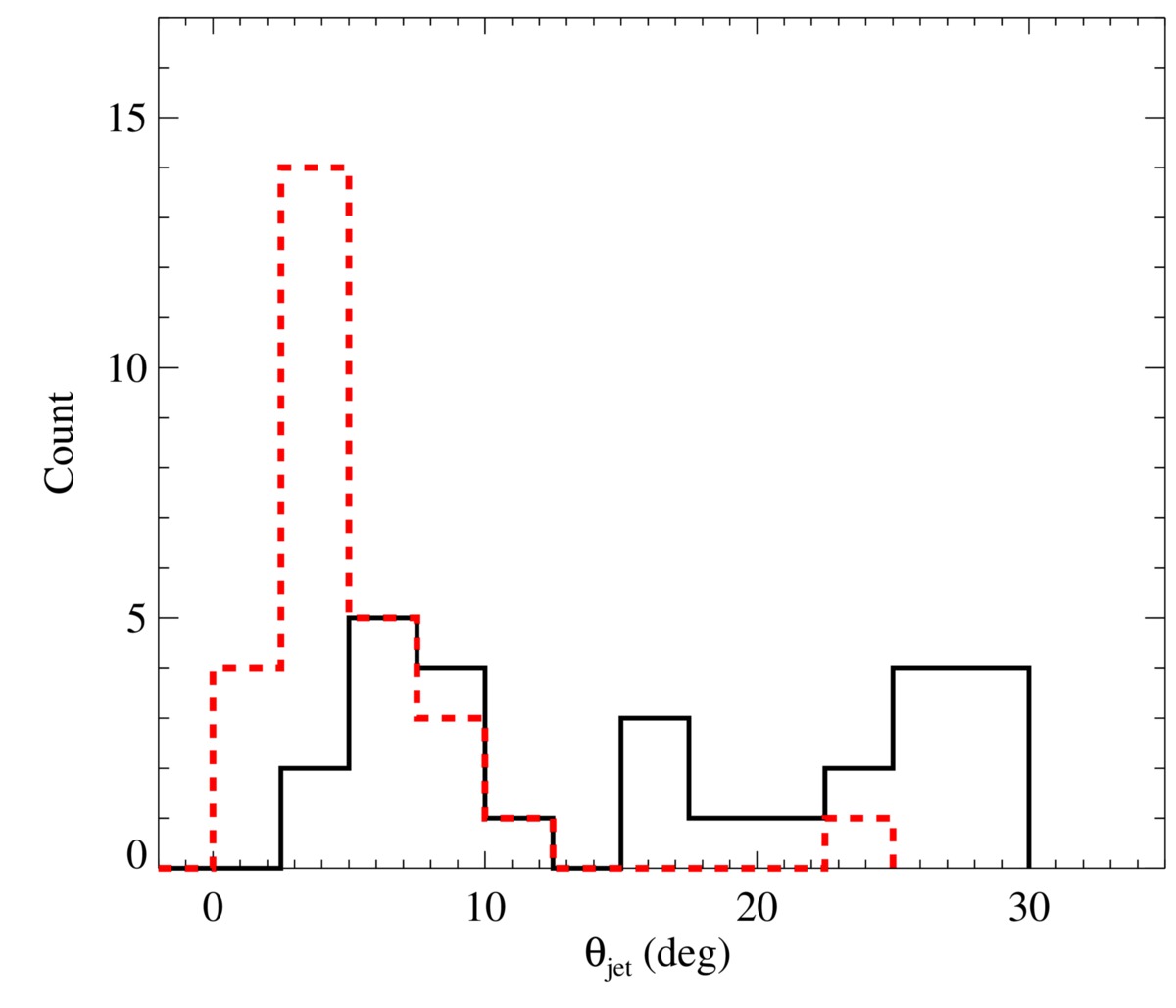}
 }
\end{center}
 \caption{%
(a) Best-fit jet opening angle, $\theta_{jet}$, for each burst. Numbers on the \textit{x}-axis (e.g, 1-27) correspond to the second column of Table~\ref{tab:main}, numbered sequentially from GRB~051121 to GRB~121027A. (b) Black line shows the histogram of the best-fit $\theta_{jet}$ values without considering their uncertainties. For comparison, we also plot the histogram of $\theta_{jet}$ values in the ``prominent" sample of Racusin et al 2009 with red dashed lines.}
\label{fig:jet_hist}
\end{figure}

\paragraph{Jet opening angles} 
Previous studies were based on the jet-break identification to constrain the jet opening angle. Missing a break in the light curve would be problematic for this method and only a lower limit of the jet opening angle could be given in such cases \citep[e.g.,][]{Racusin09}. Our numerical model, on the other hand, does not require a distinct break to be apparent in the light curve. As long as our model can fit the data, the jet-opening angle and other model parameters can be{ tested and sometimes constrained}. Furthermore, 
our technique automatically accounts for ``hidden'' jet breaks \citep{Curran08,Racusin09} by deriving confidence limits based directly on fits of the detailed hydrodynamical/radiation models to the data and{ avoids utilizing oversimplified closure relations}.

The individual fit results are shown in Figure~\ref{fig:jet_hist}a, where the points show the best-fit values and the uncertainties are derived from the distribution of numerical fit results. 
Jet opening angles are measured with typical uncertainties of order 20\% -- 50\%\footnote{We note again the uncertainties are sometimes estimated only from the ``local" mode region where the maximum likelihood is found when multi-modality happens.}. The two short bursts (\#1 and \#25 in our sample; Figure~\ref{fig:jet_hist}a) have fairly large allowed range of opening angles compared with the bulk of the long bursts, but the sample is very small. 
The values for the best-fit angles have a broad distribution (see Figure~\ref{fig:jet_hist}b). The overall conclusion is that roughly 50\% of the GRBs are consistent with opening angles in the 5-10\degree range seen in early jet break studies \citep[][]{Frail01,Bloom03,Racusin09}, but roughly 50\% are consistent with much larger opening angles, which in some cases could be larger than the limit of 28.6\degree included in our models. However we note here that our sample is subject to selection effect simply because in order to obtain late \chandra\  observations, GRBs with long-lasting X-ray light curves are more favorable than those with early jet breaks (hence small opening angles). This effect can be clearly seen by comparing the red dashed lines (the ``prominent" sample in Racusin et al 2009) with back solid lines (our sample) in Figure 2b.

\paragraph{Viewing angles} The relative off-axis angles, $\theta_{obs}/\theta_{jet}$, are shown in shown in Figure~\ref{fig:obs_angle}.  26\% of bursts (namely GRBs 051221A, 070125, 080319B, 090113, 100413A, 100816A and 110402A) have non-negligible probability of being on-axis. The rest of the bursts are consistent with, or require, large off-axis angles, and their  $\theta_{obs}/\theta_{jet}$ values have a single-peaked distribution with a peak of 0.8. This points clearly to the importance of including off-axis effects in fits to GRB afterglows.

\paragraph{Electron energy index} We found that $p$ is well constrained for most of the bursts 
(Figure~\ref{fig:p_hist}), perhaps because it is the dominant parameter determining the slopes of the light curves. Best-fit values ranged from 2.0 (the lower limit of the simulation) to 2.9. A few bursts
(including both of the short GRBs) have distributions for $p$ that are clustered close to 2.0, suggesting that even lower values might be appropriate.

\begin{figure}[ht!]
\begin{center}
\subfigure[ ]{
\includegraphics[scale=0.9]{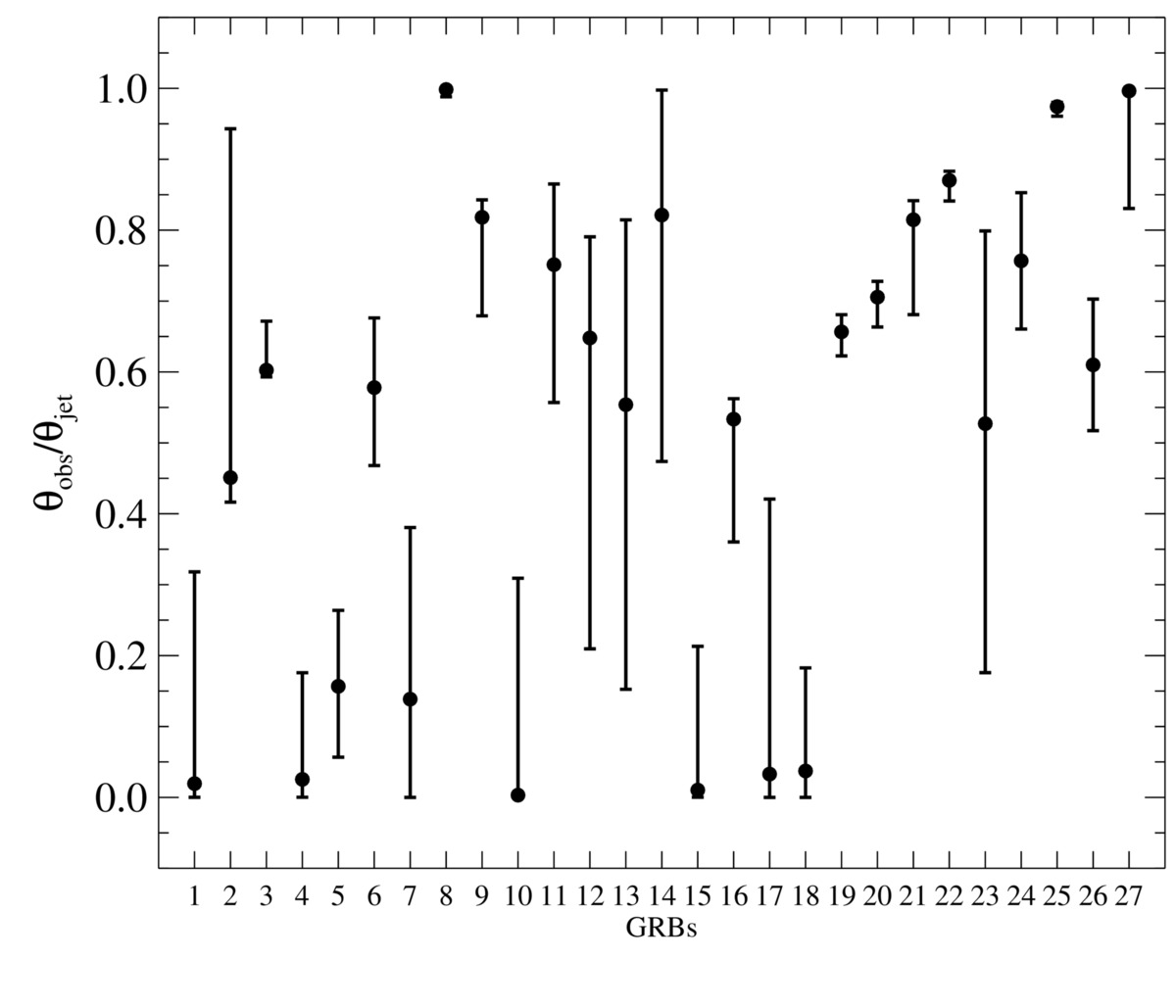}
 }
\subfigure[ ]{
\includegraphics[scale=0.9]{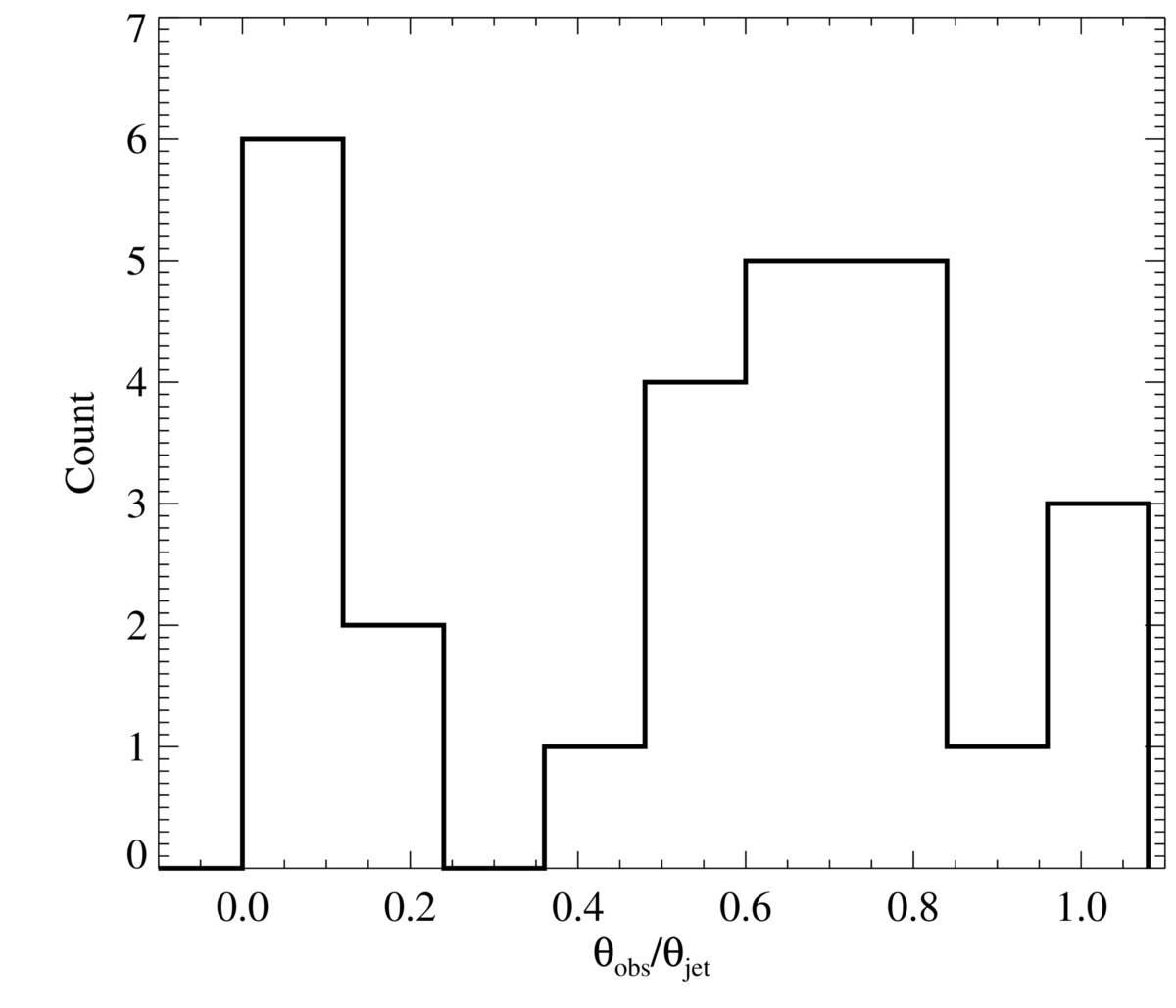}
 }
\end{center}
 \caption{%
Same as Figure~\ref{fig:jet_hist} but for $\frac{\theta_{obs}}{\theta_{jet}}$.
}
\label{fig:obs_angle}
\end{figure}

\begin{figure}[ht!]
\begin{center}
\subfigure[ ]{
\includegraphics[scale=0.9]{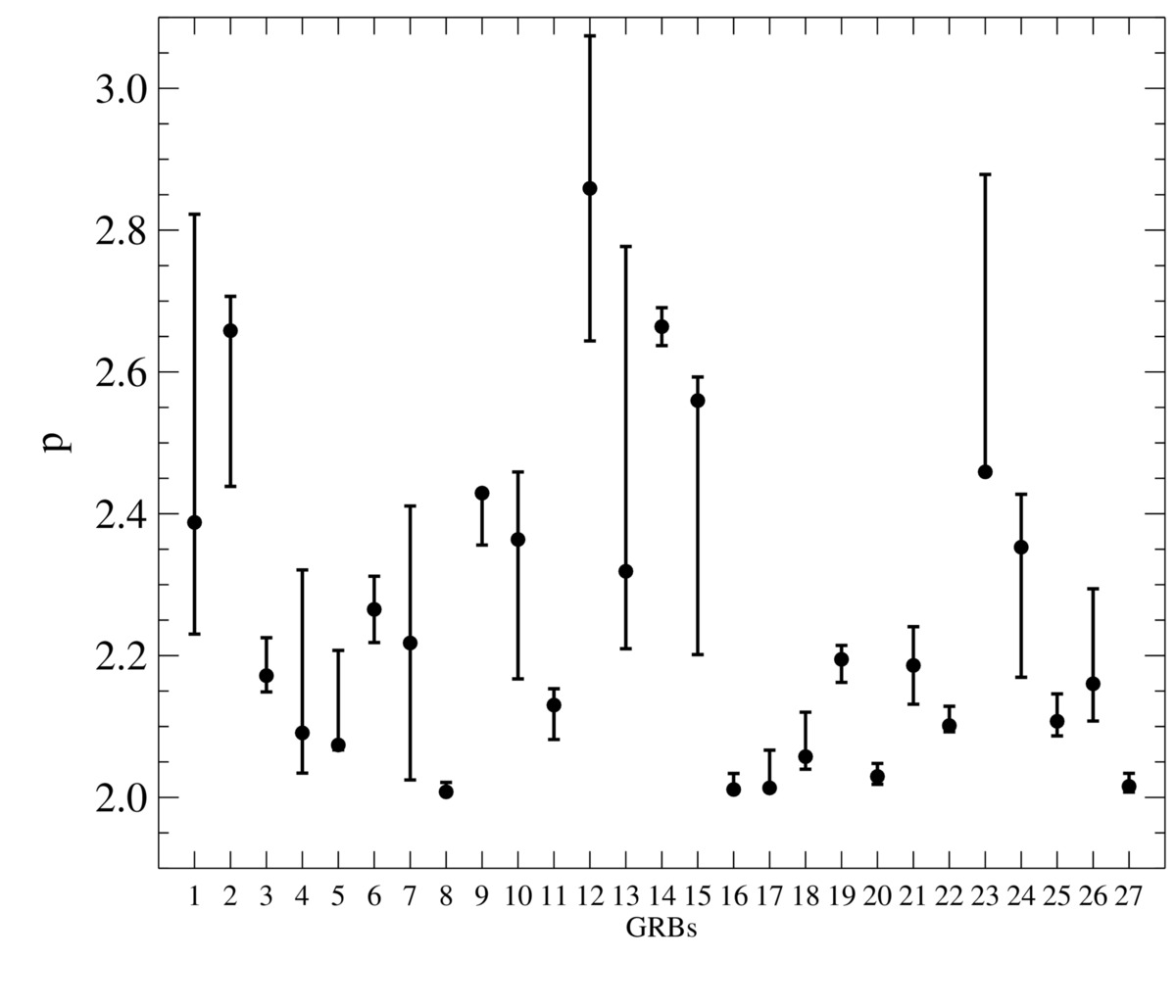}
 }
\subfigure[ ]{
\includegraphics[scale=0.9]{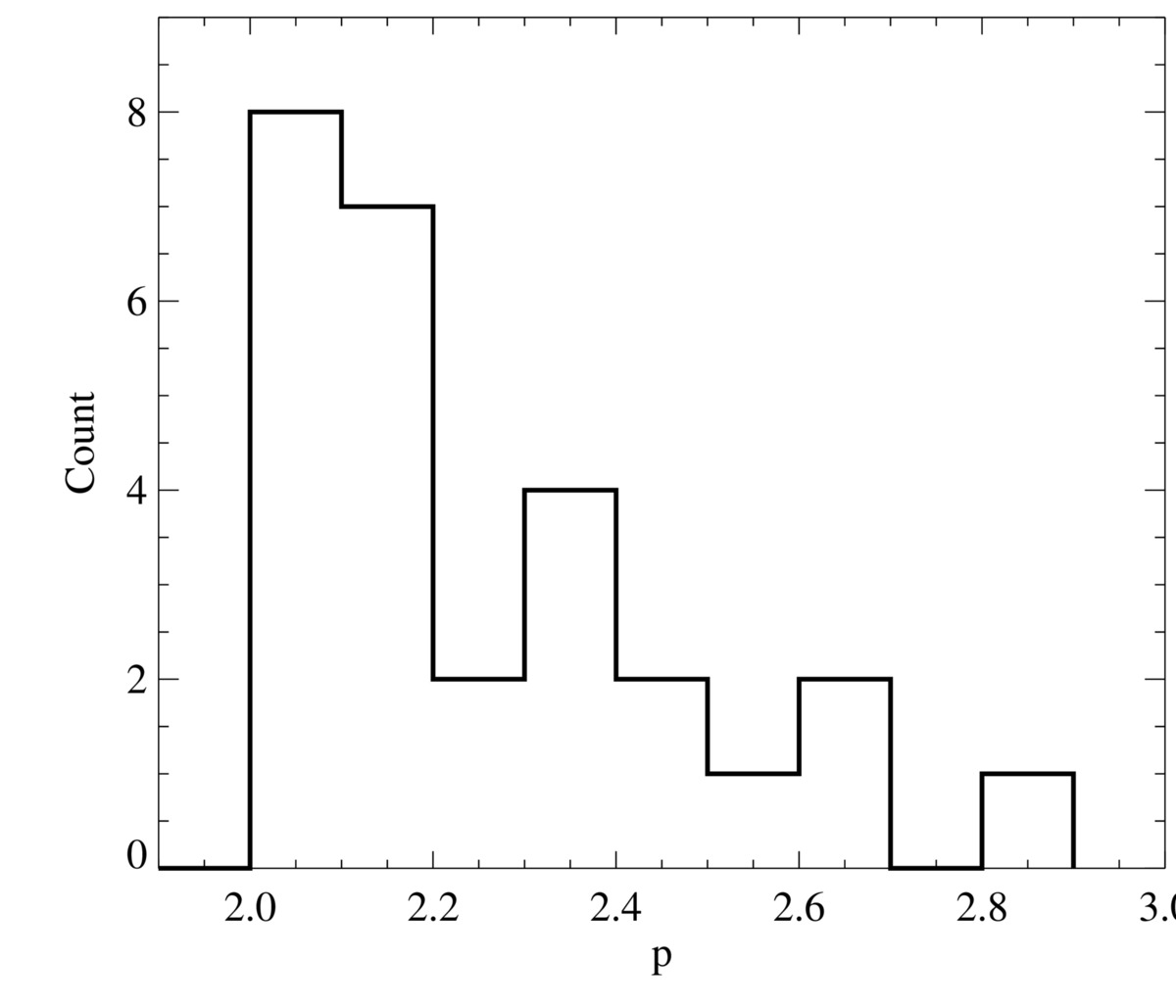}
 }

\end{center}
 \caption{%
Same as Figure~\ref{fig:jet_hist} but for $p$.
}
\label{fig:p_hist}
\end{figure}

\paragraph{Other model parameters} 
{ The distributions of $E_{53}$, $n$, $\epsilon_B$, and $\epsilon_e$ are broadly distributed due to the degeneracies in the model for data sets that do not span the synchrotron spectral breaks, $\nu_m$ and $\nu_c$. The best-fit values of these parameters often have extreme values, but always coupled with very large uncertainties that encompass more typical values. For completeness, we include the distributions for these parameters in the on-line material. Determining these parameters more precisely will require additional data from other wavelengths (optical and/or radio), which is beyond the scope of this paper.}

\section{Conclusions}

In this paper, we studied 27 well-sampled GRB X-ray afterglows with \chandra\ late and deep observations. Those data, in terms of the completeness of deepest follow-up of the afterglow, are currently the best ones in the X-ray band to address the jet physics of GRBs.

Previously studies reporting a lack of jet breaks typically apply a power law approximation to the data and simplified ``closure relations'' linking spectral and temporal shape of the observed flux. A key difference in our approach is that we directly fit models leading to complex non power-law shapes of the light curves, taking into account the following three factors: (i) lateral expansion, (ii) edge effects and (iii) jet orientation. The fluid profiles of spreading and decelerating jets have thus far not been captured analytically, and numerical simulations are needed to properly capture the blast wave dynamics. In this work, we apply directly a physical model derived from a large set of simulation-based synchrotron templates to a large set of observational data, constraining jet parameters through a nested sampling fitting method \citep[Feroz et al. 2009; see also][for a similar approach using a Monte Carlo approach to Swift/XRT data only]{2015ApJ...799....3R}. Our results benefit from the inclusion of  \chandra\ data and show good fits with collimated outflow light curves that are often observed off-axis, suggesting that the observer angle $\theta_{obs}$ must be taken into account when calculating the shape of afterglows.

There are biases in our sample due to our selection criterion of requiring late-time \chandra\ observations, which emphasizes GRBs with long-lasting X-ray light curves, and which therefore selects against bursts with early jet breaks (hence small opening angles). There may also be biases caused by the lack of models with $\theta_{jet} < 0.045$ that prevent us from finding narrow jets. 

The {\it MULTINEST} algorithm employed in this work has been extensively tested for consistency with two independently developed MCMC
parallel tempering methods \citep[see also][]{2015ApJ...799....3R}, both on samples 
with and without \chandra\ data. The tests confirm that both methods are 
capable of correctly uncovering the features of multimodal 
distributions, which is a very challenging problem for regular MCMC 
methods. We find that the model 
parameter value distributions that we obtain in this study are 
consistent with those of \cite{2015ApJ...799....3R}. The two studies provide 
complementary approaches to the jet break problem: without \chandra\ 
points, \cite{2015ApJ...799....3R} can study a relatively larger sample, while the 
current work demonstrates how the inclusion of late time \chandra\ data 
can make a strong difference for some individual cases.

There are several aspects that are not yet addressed but are important in constraining the jet physics. First, we only focused on the X-ray data; inclusion of optical and radio data would give better constraints on the physical parameters and the important cooling frequency in the afterglow spectra. Second, we did not consider the central engine contributions. We assumed that the shallow decay and flare phases of the X-ray light curves are due to the central engine activity and simply excluded them from our fits.{ A more comprehensive model should consider both central engine} and afterglow and explain the observational data in a consistent way, especially when the two contributions overlap. Third, our jet model in this work is limited to an ISM case. Other cases such as wind-blown bubbles ($n\propto r^{-2}$) or more general density distributions ($n\propto r^{-k}$) can also apply in at least some GRB cases \citep[see e.g,][]{Leventis:2013vw}. All these issues are beyond the scope of this paper and will be discussed in future works.

\acknowledgments
We thank the anonymous referee for detailed and thoughtful comments that greatly improved the paper. We thank Peter Veres, Peter M\'esz\'aros, Kazumi Kashiyama, Xiao-Hong Zhao, Xue-Wen Liu, Valerie Connaughton, Dirk Grupe, Derek Fox, Abe Falcone, Leisa Townsley, Eveline Helder, He Gao, Liang Li, Fangkun Peng, Enwei Liang, Neil Gehrels and Bing Zhang for helpful comments and suggestions. We thank Tyson Littenberg  and Dan Foreman-Mackey for discussion on the MCMC method. BBZ thanks Johannes Buchner and Farhan Feroz for help on the {\it MULTINEST} codes. This work was supported by SAO contract SV4-74018, NASA contract NAS5-00136, and by SAO grants AR3-14005X, GO1-12102X, and GO3-14067X. HVE, GSR and AM  acknowledge the support by NASA TM3-14005X and NNX13AO93G. We acknowledge the use of public 
data from the \swift~and \chandra~data archive. This work made use of data supplied by the UK Swift Science Data Centre at the University of Leicester.
{\it Facilities:} \facility{Swift} and \facility{Chandra}.

\end{document}